\documentstyle[prl,aps,twocolumn,epsfig]{revtex}

\begin{document}
\draft

\twocolumn[\hsize\textwidth\columnwidth\hsize\csname @twocolumnfalse\endcsname

\title{Spin/Orbital Pattern-Dependent Polaron Absorption in Nd$_{1-x}$Sr$_{x}$MnO$%
_3 $}
\author{M. W. Kim,$^{1,2}$ J. H. Jung,$^{1,3}$ K. H. Kim,$^{1}$ H. J. Lee,$^{1,2}$
Jaejun Yu,$^{1,3}$ T. W. Noh,$^{1,2}$ and Y. Moritomo$^{4}$}
\address{$^{1}$School of Physics, Seoul National University, Seoul 151-747, Korea}
\address{$^{2}$Research Center for Oxide Electronics, Seoul National University,\\
Seoul 151-747, Korea}
\address{$^{3}$Center for Strongly Correlated Material Research, Seoul National\\
University, Seoul 151-747, Korea}
\address{$^{4}$CIRSE, Nagoya University, Nagoya 464-8603, Japan}
\date{\today }
\maketitle

\begin{abstract}
We investigated optical properties of Nd$_{1-x}$Sr$_{x}$MnO$_{3}$ ({\it x}=
0.40, 0.50, 0.55, and 0.65) single crystals. In the spin/orbital disordered
state, their conductivity spectra look quite similar, and the strength of
the mid-infrared absorption peak is proportional to {\it x}($1-${\it x})
consistent with the polaron picture. As temperature lowers, the Nd$_{1-x}$Sr$%
_{x}$MnO$_{3}$ samples enter into various spin/orbital ordered states, whose
optical responses are quite different. These optical responses can be
explained by the spin/orbital ordering pattern-dependent polaron hopping.
\end{abstract}

\vskip1pc] \newpage

Recently, lots of attention has been paid to the roles of orbital degree of
freedom in the physical properties of manganites. Resonant x-ray scattering
of a layered manganite provided evidences of orbital ordering\cite{Murakami}%
. Polarization microscope images demonstrated the existence of optical
anisotropy \cite{Tokura science}, which might be due to the orbital
ordering. A new elementary excitation caused by symmetry breaking of orbital
ordering, called `orbiton', was proposed and claimed to be observed in LaMnO$%
_{3}$ by Raman spectroscopy measurements\cite{Saitoh}. Since the
corresponding Raman peaks were observed in the energy region of a few tenth
of eV, similar excitations could be observed by infrared spectroscopy.

Indeed, an anomalous incoherent absorption peak has been observed in the
mid-infrared (IR) region for numerous manganites\cite{Okimoto 95,Okimoto
PCMO,Ishikawa,Kim,Kaplan,Liu}. However, the origin of this mid-IR absorption
peak is not fully understood yet. In one scenario, it was suggested that the
mid-IR absorption should be due to orbital excitations\cite{Mack}. Even
though the observed anisotropy might be explained by this model, there have
been few experimental reports to compare with the theoretical predictions
even on a qualitative level\cite{Ishikawa}. In the other scenario, it has
been suggested that the mid-IR absorption should arise from hopping of the
Jahn-Teller (JT) polaron \cite{Kim,Kaplan,Liu,Millis,Yunoki}. The polaron
binding energy obtained from the mid-IR peak position agrees with that from
other measurements. However, this scenario has not been extended to explain
the observed optical anisotropy. Till now, there is no consensus on these
two scenarios. Therefore, it is important to understand the role of the
orbital degree of freedom on the mid-IR absorption peaks.

Nd$_{1-x}$Sr$_{x}$MnO$_{3}$ (NSMO) shows various types of spin/orbital (S/O)
ordering patterns at low temperature ($T$), depending on the value of the
doping concentration ({\it x})\cite{Kajimoto}. Optical investigations on
NSMO in various S/O ordered states can provide us further insights on the
role of S/O for the mid-IR absorption in manganites. In spite of such
importance, few investigations have been reported due to the difficulties in
growing high quality single-domain crystals.

In this Letter, we present optical responses of NSMO ({\it x}=0.40, 0.50,
0.55, and 0.65) multi-domain single crystals and a single-domain crystal
with {\it x}=0.55. At room $T$, optical conductivity spectra $\sigma (\omega
)$ of all the samples appear quite similar and can be explained within the
JT polaron picture. At low $T$, by taking account of the effects of the S/O
ordering pattern on the polaron hopping motion, the anisotropy in $\sigma
(\omega )$ for the single-domain sample can be explained quite well. This
S/O pattern-dependent hopping model was successfully extended to
quantitatively explain the reflectivity spectra $R(\omega )$ of other NSMO
multi-domain crystals. This clearly demonstrates that the polaron scenario
is indeed valid, and that the polaron hopping motion should be strongly
affected by the S/O ordering patterns.

The NSMO single crystals were grown by the floating zone method\cite
{Moritomo Unpublished}. After polishing and proper annealing, $R(\omega )$
were measured by near normal incident unpolarized light over a wide photon
energy region 5 meV$-$30 eV. For the single-domain {\it x}=0.55 crystal, the
crystal axes were determined using the neutron scattering measurement\cite
{Moritomo Unpublished}. Using polarized light, we obtained $R(\omega )$ for
the {\it ab}-plane, i.e. $R_{ab}(\omega )$, and for the {\it c}-axis, i.e. $%
R_{c}(\omega )$. Details of the optical measurements were described in Ref. 
\cite{Lee}.

Figure 1(a) shows $R(\omega )$ for NSMO multi-domain single crystals at room 
$T$. Note that all the samples are insulators with disordered S/O states at
room $T$\cite{Kajimoto,Kuwahara}. Clearly, $R(\omega )$ are similar for all
the samples despite the different hole concentrations. On the other hand,
Fig. 1(b) shows $R(\omega )$ for the NSMO multi-domain samples at low $T$. $%
R(\omega )$ show quite different characteristics for the samples with
different doping level {\it x}. For the {\it x}=0.40 (i.e., $F$-type
ordered) sample, $R(\omega )$ increases significantly below 1.2 eV and
approach 1.0 in the {\it dc} limit, showing metallic behavior\cite{Okimoto
95}. For the {\it x}=0.50 (i.e., $CE$-type ordered) sample, $R(\omega )$
shows an insulator-like behavior with rather strong suppression below 0.2
eV. For the {\it x}=0.55 (i.e., $A$-type ordered) sample, $R(\omega )$ keeps
increasing and approaches to about 0.6 as $\omega $ decreases.\ Finally, for
the {\it x}=0.65 (i.e., $C$-type ordered) sample, $R(\omega )$ shows an
insulating behavior. Such large differences in $R(\omega )$ at low $T$ are
not easily understood by keeping in consideration of the similarity in $%
R(\omega )$ at room $T$ and the small differences in {\it x}. It is quite
apparent\ that S/O ordering patterns could be playing an important role in
optical properties of NSMO.

We have obtained $\sigma (\omega )$ using the Kramers-Kronig (KK) analysis
of $R(\omega )$ at room $T$. Since the optical anisotropy in these
disordered S/O states is expected to be small, the KK analysis can be
applied at room $T$ even for the multi-domain crystals. As shown in Fig.
2(a), $\sigma (\omega )$ are quite similar for all the samples. In order to
compare the polaron and the orbital excitation scenarios, we should obtain
terms related to the coherent and the incoherent motions of free carriers
from $\sigma (\omega )$.

In the perovskite manganites, there are numerous optical transitions. As
shown in Fig. 2(b) for the {\it x}=0.55 sample, the room temperature $\sigma
(\omega )$ below 4.0 eV was fitted with four peaks represented by the dotted
lines. The first peak (Peak I) was due to either polaron hopping motions\cite
{Millis,Yunoki} or orbital excitations\cite{Mack}. The second peak (Peak II)
located around 1.5 eV was suggested to be due to the transition between JT
split {\it e}$_{g}$ bands\cite{Jung98}. [A Monte Carlo simulation using a
Hamiltonian with the double exchange and JT interactions also clearly
predicted the existence of Peaks I and II\cite{Yunoki}.] The third peak
(Peak III) was suggested to be due to a transition between the Hund's rule
exchange ($J\sim $ 2 eV) split bands\cite{Moritomo-2}. Finally, the fourth
peak (Peak IV) was proposed to be due to the charge transfer transition
between the O 2{\it p} and Mn 3{\it d} levels\cite{Lee}. Note that the
position of each peak are quite consistent with the results of other
measurements\cite{Jaime}, and the validity of those peaks has been suggested
by many groups\cite{Liu,Lee,Quijada} and such analyses have already been
performed quite successfully to explain the optical responses of (Bi,Ca)MnO$%
_{3}$\cite{Liu}, Nd$_{0.7}$Sr$_{0.3}$MnO$_{3}$\cite{Lee}, and (La,Ca)MnO$%
_{3} $\cite{Jung98}.

In the polaron scenario, Peak I should be due to the JT polaron hopping
motion. Since it can be considered as the interatomic transition from the JT
distorted Mn$^{3+}$[{\it x} ions per unit volume] to Mn$^{4+}$[(1$-${\it x})
ions per unit volume], it should be proportional to {\it x}(1$-${\it x}).
The inset in Fig. 2(a) shows the {\it x}-dependent strength of Peak I ($%
S_{I} $). The dotted line in the inset shows, within our fitting error bars,
that the {\it x}-dependence of $S_{I}$ is well fitted with {\it x}(1$-${\it x%
}). A similar behavior was also observed in (La,Ca)MnO$_{3}$\cite{Jung98},
supporting the agreement with the polaron scenario. However, as far as we
know, the {\it x}-dependence of Peak I in the orbital scenario, i.e., the
incoherent motion of free carriers due to orbital excitations, has not been
predicted yet.

To obtain further insights, we have investigated $\sigma (\omega )$ for the 
{\it x}=0.55 {\it single-domain} crystal at low $T$ in the $A$-type S/O
ordered state. The open triangles in Figs. 3(a) and 3(b) show $\sigma
(\omega )$ along the {\it c}-axis, $\sigma _{c}(\omega )$, and in the {\it ab%
}-plane, $\sigma _{ab}(\omega )$, respectively. Clearly, there exists a
rather strong anisotropy of $\sigma (\omega )$ below 1.5 eV. The solid lines
show the fitting curves based on the above-mentioned mode assignments. As
shown in the dotted lines, the strengths of Peaks II and IV were assumed to
be $T$-independent, and the strength of Peak III were assumed to be nearly
zero. Since Peaks II and IV are due to the interband transitions between JT
split {\it e}$_{g}$ bands and the charge transfer transition between the O 2%
{\it p} and Mn 3{\it d} bands, respectively, they should be nearly $T$%
-independent. And, since Peak III is due to a transition between the Hund's
rule exchange ($J\sim $ 2 eV) split bands\cite{Moritomo-2}, it should vanish
in the spin ordered states. On the other hand, the strength of Peak I is
suppressed along the {\it c}-axis but increases in the {\it ab}-plane. The
enhancement factor of Peak I in the $A$-type S/O ordered state along the 
{\it c}-axis, $\eta _{A}^{c}$, was estimated to be about 0.4, and that in
the {\it ab}-plane, $\eta _{A}^{ab}$, about 1.8. The difference between $%
\eta _{A}^{c}$ and $\eta _{A}^{ab}$ for Peak I can explain the reported
optical anisotropy in the S/O ordered state. It is important to note that
all the parameters, except the enhancement factors, are fixed in the fitting
process: namely, the solid lines in Figs. 3(a) and 3(b) are the results of
single parameter fittings, keeping all other parameters the same as in the
room temperature $\sigma (\omega )$ fitting.

How can we explain the difference between $\eta _{A}^{c}$ and $\eta
_{A}^{ab} $? Recently, $\sigma (\omega )$ for the planar orbital ordered
phase of the {\it e}$_{g}$ electron with {\it x}$^{2}-${\it y}$^{2}$
symmetry (the $A$-type orbital ordered state)\ were calculated based on the
orbital excitation scenario\cite{Mack}. This 2-dimensional model predicted
that the spectral weight of the incoherent absorption, $S_{I}$, should
decrease with decreasing $T$. On the other hand, that of the coherent
absorption, corresponding to the Drude peak, should increase. Contrary to
the predictions, the experimental value of $\eta _{A}^{ab}$ becomes larger
than 1.0. Moreover, Fig. 3(b) does not show any significant Drude-like
absorption term. [Note that the value of $\sigma _{ab}(\omega )$ at the {\it %
dc} limit in Fig. 3(b) was estimated at around 2$\times $10$^{2}$ $\Omega
^{-1}cm^{-1}$, which is consistent with the transport experiment value on a
single-domain crystal \cite{Kuwahara}]. Therefore, the model based on the
orbital excitation scenario cannot explain our experimental data.

As far as we know, there are no explicit predictions on how the S/O ordering
pattern will affect the polaron absorption. However, as shown in Fig. 3(c),
it is quite clear that the polaron hopping motion along the {\it c}-axis
should be suppressed since the antiparallel spin alignment in the $A$-type
S/O ordered state will cause a large energy cost of $J$. Due to the {\it x}$%
^{2}-${\it y}$^{2}$ {\it e}$_{g}$ orbital ordering within the {\it ab}%
-plane, the overlap of two Mn {\it e}$_{g}$ orbitals via an oxygen {\it p}$%
_{z}$ orbital along the {\it c}-axis becomes small, thereby suppressing $%
S_{I}$ even further. On the contrary, $S_{I}$ in the {\it ab}-plane will
become enhanced due to the ferromagnetic spin alignment and the increase of
orbital overlap in the {\it x}$^{2}-${\it y}$^{2}$ orbital ordered state.

In the spin/orbital pattern dependent polaron (SOPDP) model, the changes in $%
S_{I}$ in the S/O ordered state should be calculated by considering the
orbital contributions in the conventional polaron models, which usually
include the double exchange and the JT polaron terms. However, this
calculation might be quite difficult. In this paper, we made simple and
crude estimations of the enhancement factors using the matrix element terms
in the dipole transition. First, we considered the JT polaron absorption as
a transition from a lower {\it e}$_{g}$ level of a JT distorted Mn$^{3+}$
site to one of doubly degenerate {\it e}$_{g}$ levels of Mn$^{4+}$ sites.
Second, for the $A$-type ordered state, we evaluated the sum of possible
dipole matrix elements considering the neighboring S/O configurations. For
the Mn 3{\it d} {\it e}$_{g}$ and the O 2{\it p }orbital wave functions, we
used the wave functions of isolated ions located at corresponding sites.
Third, for the disordered state, we evaluated the sum of possible dipole
matrix elements, assuming that there is no preference of S/O for the initial
state. Finally, we divided the summed values to obtain enhancement factors
of $S_{I}$ in the ordered state. Details of this estimation procedure will
be published elsewhere\cite{MWKim}. We found that $\eta _{A}^{c}\sim $ 0 and 
$\eta _{A}^{ab}\sim $ 1.7. The estimated value of $\eta _{A}^{ab}$ agrees
quite well with the experimental value, suggesting the validity of the
SOPDP\ model. The suppression of $S_{I}$ along the {\it c}-axis can be
explained at least qualitatively. However, the difference between the
experimental and the estimated values of $\eta _{A}^{c}$ might be due to the
experimental errors and/or the over-simplification of our estimation.

For the other multi-domain single crystal samples, the polarized optical
microscope images showed that they consist of multi-domains whose typical
size is a little larger than 10 $\mu m$. Thus, the KK analysis based on $%
R(\omega )$ for multi-domain crystals may lead to errors in the positions
and strengths of Peaks I and II \cite{MWKim}. To avoid such errors, we
decided to analyze our $R(\omega )$ directly. The optical responses of
inhomogeneous media have been investigated extensively. In the long
wavelength limit, where the wavelength $\lambda $ of the incoming light is
larger than the domain size $L$, the medium should look homogeneous with an
effective conductivity\cite{Noh}. On the other hand, if $\lambda <L$, the
light will see the responses of individual domains, so we will obtain
averaged optical responses. If the {\it c}-axes of the domains are randomly
oriented along any one of the crystallographic axes, the measured $R(\omega
) $ is given by [$\frac{2}{3}R_{ab}(\omega )$+$\frac{1}{3}R_{c}(\omega )$].

To test the above argument, we have calculated $R(\omega )$ for the
multi-domain case of $A$-type (i.e., {\it x}=0.55) ordering, whose
polarization-dependent $\sigma (\omega )$ were obtained experimentally. As
mentioned above, the predictions of the SOPDP model are quite well matching
to the real $\sigma (\omega )$ of a single-domain crystal. From $\sigma
_{ab}(\omega )$ and $\sigma _{c}(\omega )$ of the SOPDP model, as shown in
Figs. 3(a) and 3(b) (solid lines), respectively, corresponding reflectance
spectra for given polarization directions were obtained. Using the predicted 
$R_{ab}(\omega )$ and $R_{c}(\omega )$, $R(\omega )$ were evaluated and is
shown as the open triangles in Fig. 1(b). The excellent agreement between
the experimental data and the SOPDP model predictions suggests that the
method of obtaining the $R(\omega )$ for a multi-domain sample using the
SOPDP model is accurate for $\lambda <L\sim 10$ $\mu m$.

For the $F$-type (i.e., {\it x}=0.40) sample, which has isotropic S/O
structure even at low $T$, the enhancement of the mid-IR region and the
metallic behavior can be explained well with the existing JT polaron model,
which considers the JT distortion and double exchange interaction, as
observed in La$_{0.7}$Ca$_{0.3}$MnO$_{3}$\cite{Kim}. For other multi-domain
samples, similar analyses based on the SOPDP model were also performed\cite
{MWKim}. For the $C$-type ordered (i.e., {\it x}=0.65) sample, $\eta
_{C}^{c} $ $\sim $ 2.9 and $\eta _{C}^{ab}$ $\sim $ 0 can provide a good
description to our experimental data. [Here we used the same method with $%
\eta _{A}$ \ using {\it 3z}$^{2}${\it -r}$^{2}$ orbital for the initial
state instead of {\it x}$^{2}${\it -y}$^{2}$ at low {\it T}.] For the $CE$%
-type (i.e., {\it x}=0.50) sample , this approach can provide a reasonable
fit to the experimental $R(\omega )$ when Peak I becomes asymmetric\cite
{MWKim}. It might be due to the opening of the charge gap at the low
frequency region in the S/O ordered state. As shown in Fig. 1(b), most
experimental $R(\omega )$ can be explained by the theoretical predictions
using the SOPDP model with a single fitting parameter ($\eta $). The good
agreements suggest that the {\it optical response observed in various
spin/orbital ordered states of NSMO can be explained quite well within the
SOPDP model}.

In conclusion, we have presented the temperature-dependent spectral weight
changes of Nd$_{1-x}$Sr$_{x}$MnO$_{3}$ single crystals, which strongly
depend on the pattern of the spin/orbital ordering at low temperature. These
optical responses can be described quantitatively by taking account of
spin/orbital pattern-dependent polaron hopping.

This work was financially supported by the Ministry of Science and
Technology through the Creative Research Initiative Program. The work by
Y.M. was supported by a Grant-In-Aid for Scientific Research from the
Ministry of Education, Science, Sports, and Culture, and from PRESTO, JST.
KHK is also supported by the BK-21 Project of the Ministry of Education.

\begin{figure}[tbp]
\epsfig{file=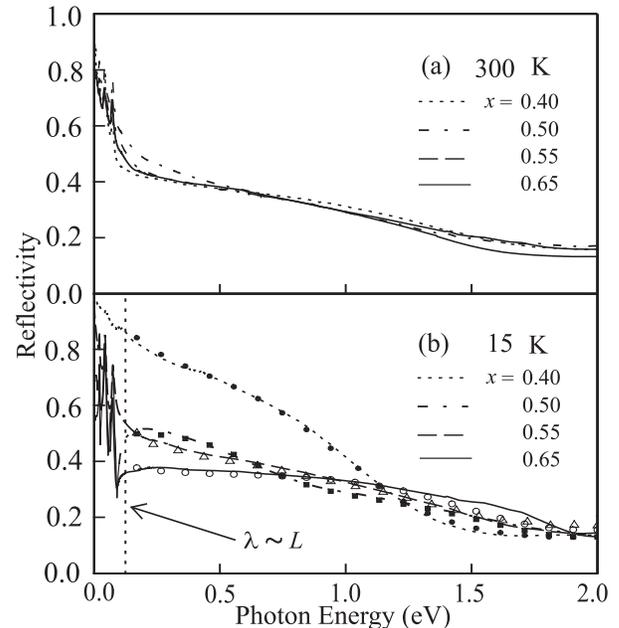,width=3.1in}
\caption{$R(\protect\omega )$ of NSMO at (a) 300 K and (b) 15 K. In (b),
solid circles, solid squares, open triangles, and open circles are the
predictions of the SOPDP model for {\it x}=0.40, 0.50, 0.55, and 0.65,
respectively.}
\label{Fig:1}
\end{figure}

\begin{figure}[tbp]
\epsfig{file=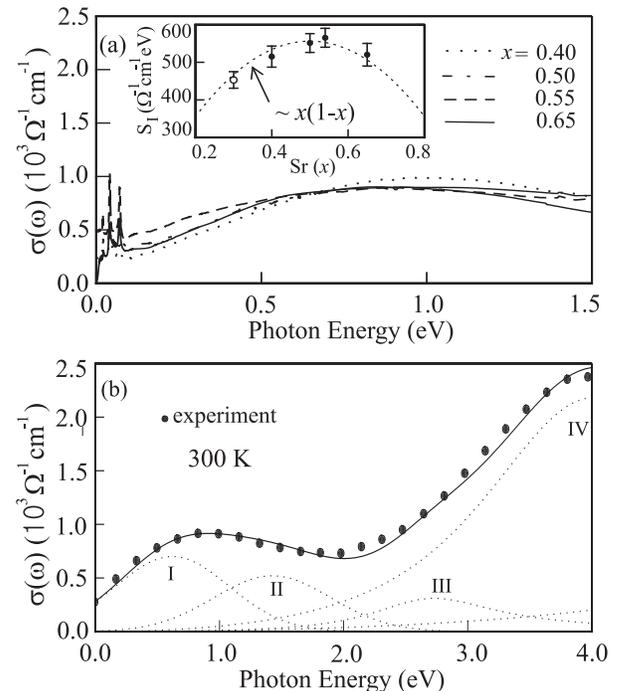,width=3.1in}
\caption{(a) $\protect\sigma (\protect\omega )$ of NSMO at room $T$. The
inset shows {\it x-}dependent $S_{I}$. The solid circles represent data from
this study, and the open circle is from the published data by Lee {\it et al}
\protect\cite{Lee}. The dotted line in the inset represents the functional
form of {\it x}(1$-${\it x}). (b) $\protect\sigma (\protect\omega )$ of NSMO
({\it x}=0.55) at room $T$ (solid circles), where the solid line and dotted
lines represent the fitting curves.}
\label{Fig:2}
\end{figure}

\begin{figure}[tbp]
\epsfig{file=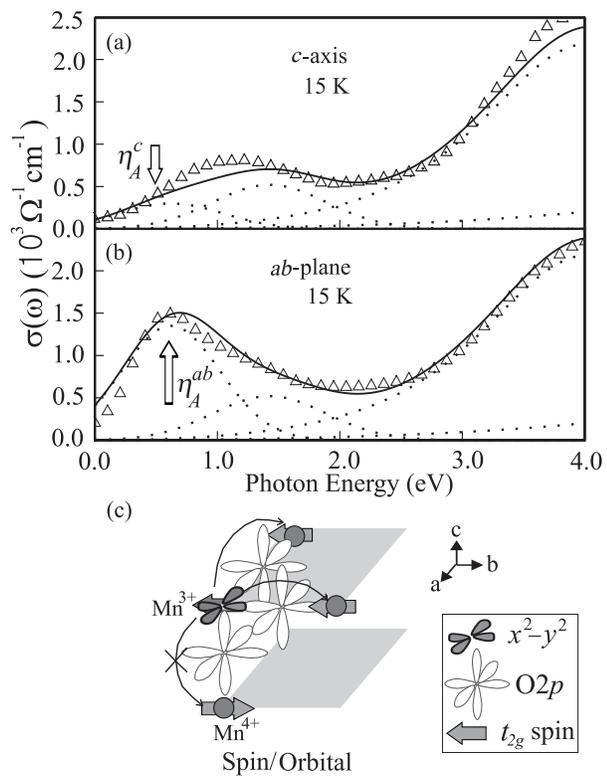,width=3.1in}
\caption{The experimental (open triangle) and SOPDP model-predicted $\protect%
\sigma (\protect\omega )$ (solid line) of NSMO ({\it x}=0.55) at low $T$ for
(a) the {\it ab}-plane and (b) the {\it c}-axis. (c) Schematic
representation of the SOPDP model in the $A$-type ordered state.}
\label{Fig:3}
\end{figure}

\end{document}